\documentclass[twocolumn,showpacs,preprintnumbers,amsmath,amssymb]{revtex4}

\usepackage{graphicx}
\usepackage{dcolumn}
\usepackage{bm}

\begin{document}

\title{Multispeckle diffusing-wave spectroscopy: \\
a tool to study slow relaxation and time-dependent dynamics}

\author{Virgile Viasnoff}
\affiliation{
L.P.M UMR7615 CNRS ESPCI 10 rue Vauquelin 75231 Paris, France
}
\author{Fran\c{c}ois Lequeux}
\affiliation{
L.P.M UMR7615 CNRS ESPCI 10 rue Vauquelin 75231 Paris, France
}
\author{D. J. Pine}
\affiliation{
Departments of Chemical Engineering and Materials, University of California, Santa Barbara, CA  93106-5080
}
\date{\today}
\begin{abstract}
 A multispeckle technique for efficiently measuring correctly ensemble-averaged intensity autocorrelation functions of scattered light from non-ergodic and/or non-stationary systems is described.
 The method employs a CCD camera as a multispeckle light detector and a computer-based correlator, and permits the simultaneous calculation of up to 500 correlation functions, where each correlation function is started at a different time.
 The correlation functions are calculated in real time and are referenced to a unique starting time.
 The multispeckle nature of the CCD camera detector means that a true ensemble average is calculated; no time averaging is necessary.
 The technique thus provides a ``snapshot" of the dynamics, making it particularly useful for non-stationary systems where the dynamics are changing with time.
 Delay times spanning the range from 1~ms to 1000~s are readily achieved with this method.
 The technique is demonstrated in the multiple scattering limit where diffusing-wave spectroscopy theory applies.
 The technique can also be combined with a recently-developed two-cell technique that can measure faster decay times.
 The combined technique can measure delay times from 10~ns to 1000~s.
 The method is peculiarly well suited for studying aging processes in soft glassy materials, which exhibit both short and long relaxation times, non-ergodic dynamics, and slowly-evolving transient behavior.
\end{abstract}


\maketitle

\section{Introduction}

 Dynamic light scattering has proven to be a powerful tool for investigating
the dynamics of colloidal suspensions, emulsions, foams, and other complex
fluids.
 Over the past decade, dynamic light scattering (DLS) has been increasingly applied to glassy systems with slow, and often non-ergodic,
dynamics.
 Slow dynamics pose experimental difficulties for DLS, particularly with respect to the stability of the lasers, and sometimes for the light
detection system.
 Nonergodic samples pose even more serious experimental difficulties, which are exacerbated by the presence of slow dynamics.
 One way of addressing the problem of long data acquisition times is to use an
array detector.
 More importantly, an array detector can also address the
problem of non-ergodic samples if the optics are designed so that the
different elements in the array detector constitute essentially
statistically independent samplings of the system under study.

In this article, we describe a method which employs a charge-coupled device
(CCD) array detector for performing dynamic light scattering on systems that
exhibit strong multiple light scattering. While the basic method we describe
is applicable to single as well as multiple light scattering, our focus here
is on multiple dynamic light scattering, also known as diffusing-wave
spectroscopy (DWS)~\cite{maret87,pine88,weitz93}, where the methods we describe are most effective.

\section{Background}

In conventional DWS (or DLS), the basic idea is to illuminate a sample with
a coherent beam of light from a laser and to measure the temporal
fluctuations in the resulting speckle pattern of scattered light. In a
typical experiment, the optics for collecting the scattered light is
arranged in such a way that the intensity of a single speckle is measured.
By focusing on a single speckle, the amplitude of the fluctuating intensity
is optimized to obtain a strongly fluctuating signal. The temporal duration
of a typical fluctuation, \textit{i.e.}\ the lifetime of a speckle, is
determined by the time it takes a particle to move such that the phase of
the scattered light is changed by approximately unity (or $\pi$). The
dynamics of the scattered light, and hence the dynamics of the scatterers,
can be more quantitatively characterized by the temporal autocorrelation
function of the scattered intensity. In many cases, the autocorrelation
function can be analyzed so that the time dependent mean square displacement
of the scatterers is extracted~\cite{weitz93}.

There are two important keys to the success of the conventional analysis that we wish
to focus on here. First, the process must be stationary, at least for the
duration of the experiment. That is, the dynamics of the system cannot
change significantly during this time period. For example, the gelation process 
of a colloidal suspension can be monitored provided that the
particle dynamics slow down (due to the increase of the elastic modulus) occurs on time scales
longer than the measurement period. In general an accurate measurement
of the correlation function requires that the fluctuations in the intensity
be measured over many fluctuation periods (\textit{e.g.}\ $\sim 10^4$
speckle lifetimes to achieve $\sim 1$\% accuracy). Thus, in this case, the
particle dynamics must not change significantly over a time period of $\sim
10^4$ speckle lifetimes.

Second, the system must be ergodic. For a light scattering experiment, a
system is ergodic if each speckle samples the same distribution of light
intensities as the entire ensemble of speckles over the period of time that
data is acquired (\textit{i.e.}\ over the duration of the experiment). While
this condition is satisfied for a large number of systems, there are certain
systems with very slow glassy dynamics that are of interest but which,
nevertheless, are not ergodic in the sense described above. Such systems
pose a problem for light scattering and special steps must be taken in order
to assure that meaningful correlation functions, which represent dynamics
characteristic of the entire system, are measured~\cite{pusey89,joosten90,xue92}.

The problem for light scattering presented by nonergodic samples was first
understood in a clear fashion by Pusey and van~Megen who, at the same time,
presented a clever method of obtaining correctly averaged temporal
autocorrelation functions~\cite{pusey89}. This technique has proven to be
very useful but has some drawbacks: it sacrifices some signal-to-noise in
order to obtain a useful distribution of light intensities, and the dynamics
of the system must not change significantly over the duration of the
experiment~\cite{pusey89,joosten90,xue92}. Moreover, it suffers from the same
limitation as normal DLS experiments in that data must be collected over a
time scale that is much longer that the longest time scale of the
fluctuations one is trying to measure.

Other schemes have been developed which overcome some of these difficulties,
but not without a significant cost.
 The sample can be translated or rotated, for example, so that an ensemble average is obtained by brute force~\cite {xue92}.
 However, the rotation or translation introduces an artificial characteristic time scale into the correlation function, namely the time scale over which the system samples new speckles by virtue of the motion.
 Care must be taken to assure that this
time scale is much longer than any dynamical time scale of the system one is
interested in measuring.
 And once again, the duration of the experiment must be such that averaging occurs over a time scale that is long compared to the time scale introduced by the sample motion (this amounts to assuring that a sufficiently large number of speckles are sampled -- \textit{e.g.}\ $\sim 10^4$ -- to obtain the desired level of accuracy -- \textit{e.g.}\ $\sim 1\%$).
 Another method, developed recently by Scheffold \textit{et al.}~\cite{skipetrov96,romer00,scheffold01}, uses two cells, one containing the sample with the slow nonergodic dynamics that is being studied, and another containing diffusing particles with ergodic dynamics, which serves to spatially average the nonergodic signal.
 This technique is only useful for experiments involving strong multiple scattering, but nevertheless provides a simple and efficient means of dealing with nonergodic systems.
 However, the only time scales probed in the primary cell are those that are shorter than the diffusive time scale set by the secondary cell.

An attractive alternative to the techniques described above for dealing with
nonergodic samples is to use a CCD array, conveniently contained in a
commercial CCD camera, as a light detector~\cite{wiltzius93, sillescu97,
tsui98, cipelletti99, cipelletti00, knaebel00}. This allows the intensity
from a large number of speckles to be simultaneously and independently
measured. A time series of images of scattered light obtained from the CCD
camera are stored in a computer and processed such that each pixel in the
CCD array serves as an input to its own correlator. Since the fluctuation of
each speckle is independent of the other speckles, processing scattered
intensity data collected with a CCD camera is like have $N$ independent
samplings of the data, where $N$ is the number of speckles in the field of
the CCD array. If $N$ is sufficiently large, \textit{e.g.}\ $N \sim 10^4$,
then a complete ensemble average is obtained without any need to sample the
system over times any longer than the actual dynamical time scales that are
under study. Thus, a direct ensemble average can be obtained without any
need for time-averaging whatsoever. This approach solves several problems at
once. First, it solves the nonergodicity problem because a full ensemble
average (of $\sim 10^4$ samples) is obtained. Second, the process under
study need not be a stationary process because no time averaging is
necessary, provided a sufficiently large number of speckles are used. Thus,
an essentially instantaneous snapshot of the system dynamics can be taken.
If those dynamics change in time, they can be followed simply by
measuring temporal autocorrelation functions at different starting times.

There are two practical limitations of the CCD-camera-based technique.
First, the minimum delay time for which the correlation function can be
measured is set by the frame rate of the camera. In the experiments
described in this article, we use a frame rate of 500 frames per second.
Thus, delay times as small as a few milliseconds are accessible. It is
interesting and important to note that real time processing of the images is
necessary since, to our knowledge, no commercial buffer memory is able to
store them for the entire sequence of frames obtained in the course of a single experiment, which is typically on the order of $10^5$ to $10^7$ frames.
 Recent advances in microprocessor speed allows real time computation of the correlation function.
 Hence the limiting factor for frame rate is the PCI bus velocity for transferring data from the acquisition board to the computer RAM.

The second limitation of the technique is the limited sensitivity of the CCD
array used, as compared to that of a photomultiplier. This means that
relatively strong scattered light signals are needed. In practice a power of 400 mW gives a satisfactory signal to noise ratio for a frame rate of 500 per seconds. 

In this article we first describe the CCD optical set up and calculational
scheme for our computer-based correlator. In the next section we explain how
a single camera can be used to take up to 500 correlation functions
simultaneously, with the reference starting time shifted by different
amounts. We then show how the CCD camera technique can be coupled with the
double cell technique to extend the time window over which measurements can
be made from 10$^{-8}$~s to 10$^{4}$~s.

\section{Experimental Setup}

\begin{figure}[tbp]
\begin{center}
\includegraphics[width=8cm,clip]{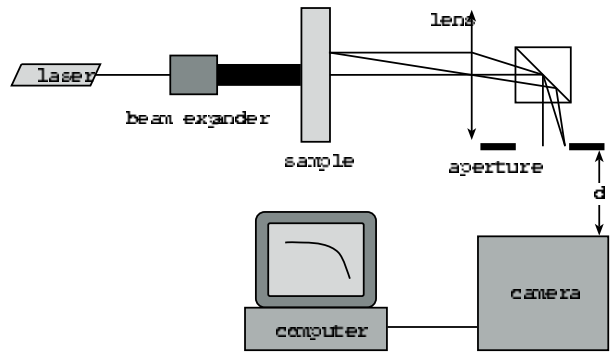}
\end{center}
\caption{Experimental set up for the transmission geometry. The laser
wavelength is $\lambda =514$~nm. We use a Dalsa Camera running at 500
frames/sec and a 500-MHz PC.}
\label{fig:dessinmontage}
\end{figure}

The sample is illuminated using polarized light from an argon-ion laser
operating at a wavelength of $\lambda = 514$~nm. Its maximal output power is 1.6 Watts. The laser beam is expanded
to a diameter of approximately 1~cm and is incident on the sample cell, as
shown in Fig.\ \ref{fig:dessinmontage}. Multiply scattered light is
collected by a 50-mm Nikon camera lens and focused onto a iris diaphragm.
The lens is set up so that a one-to-one image of the scattered light from
the output plane of the sample cell is formed on the diaphragm. The imaging
of the output plane of the sample onto the pinhole provides a convenient
means for collecting data from different parts of the sample. This feature
is particularly useful in samples like gels or very viscous solutions where
defects or small particles of dust may hinder the measurement; this set-up
allows one to avoid defected regions of the sample. A polarizing cube is
placed between the lens and the diaphragm to protect the camera from stray
light.

The low sensitivity of CCD arrays relative to photomultiplier tubes means
that the sample must be strongly illuminated. However, the incident light
intensity must not be so high as to heat the sample. Therefore, it is
crucial to collect as much scattered light as possible. We use a camera lens
of 50~mm with a numerical aperture of 1.7.

The scattered light is detected by a Dalsa CCD camera, model CAD1-128A,
which is placed approximately $d\simeq 15$~cm behind the diaphragm. It is an
8-bit camera which we run at 500 frames per second. The images are
transferred to a computer running at 500~MHz using a National Instruments
data acquisition board, model PCI1422. The combined speeds of the computer
and data acquisition boards are sufficiently high to allow data to be
analyzed in real time.

Data analysis includes calculation of the ensemble-averaged temporal
correlation function of the scattering intensity $I$, 
\begin{equation}
g_{2}(t,t_{0})\equiv \frac{\langle I(t_{0})I(t_{0}+t)\rangle }{\langle
I(t_{0})\rangle \langle I(t_{0}+t)\rangle }\;,  \label{eq:g2def}
\end{equation}
and the ensemble-averaged spatial correlation function 
\begin{equation}
C(\delta \mathbf{r})\equiv \frac{\langle I(\mathbf{r})\;I(\mathbf{r}+\delta 
\mathbf{r})\rangle }{\langle I(\mathbf{r})\rangle ^{2}}
\end{equation}
where $t_0$ is the time when the reference intensity is measured, $t \ge t_0$%
) is the delay time, $\mathbf{r}$ is the coordinate of a pixel on the
detector, and $\mathbf{r}+\delta \mathbf{r}$ is the coordinate of another
pixel a distance $\delta \mathbf{r}$ away.<> represent the average over the CCD chip. The denominator of the expression
on the right hand side of Eq.\ (\ref{eq:g2def}) is the square of the average
intensity determined according to the method suggested by Schatzel ~\cite{schatzel91}, in order
to obtain a slightly more accurate measurement of $g_{2}(t,t_0)$. The
algorithms used to calculate these correlation functions are described in
greater detail below.

\subsection{Speckle size adjustment}

\label{sec:spklsz}

The size $s$ of the speckles of scattered light formed on the CCD chip is
given by 
\begin{equation}
s \simeq d\lambda/a
\end{equation}
where $d$ is the distance between the diaphragm and the CCD chip, $\lambda$
is the wavelength of light, and $a$ is the diameter of the diaphragm. By
adjusting $d$ and/or $a$, the size of the speckles relative to the size of a
pixel on the CCD chip can be controlled. We use a 2~mm $\times$ 2~mm CCD
chip with $128\times 128$ pixels. Each pixel occupies approximately $256~\mu 
\text{m}^{2}$ with a 100\% fill factor.

To maximize the signal-to-noise, we need to collect data from as many
speckles of scattered light as possible. However, if the speckles are too
small, for example if several speckles occupy a single pixel, optical
contrast is lost and the signal for determining a temporal autocorrelation
function $g_{2}(t)$ is degraded. Therefore, it is important to adjust the
speckle size so that we obtain a satisfactory optical contrast with as much
speckles as possible analyzed.

\begin{figure}[tbp]
\begin{center}
\includegraphics[width=8cm,clip]{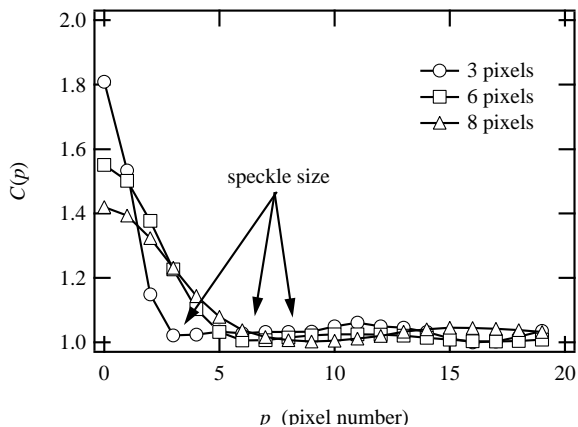}
\end{center}
\caption{Spatial correlation function $C(p)$ obtained using different
speckle sizes. The first minimum of this function is taken as the speckle
spot diameter.}
\label{fig:corpointnvnorm}
\end{figure}

In order to determine the number of speckles per pixel, we adjust the
collection optics (\textit{i.e.}\ the size of the iris) and measure the
spatial intensity autocorrelation function 
\begin{equation}
C(p)=\frac{\langle I_{i}\;I_{i+p}\rangle _{i}}{\langle I_{i}^{2}\rangle _{i}}%
\;,
\end{equation}
where $I_{i}$ is the intensity of scattered light at pixel $i$ and $I_{i+p}$
is the intensity of light $p$ pixels away. The correlation function $C(p)$
is determined by averaging over all pixels. Figure \ref{fig:corpointnvnorm}
shows $C(p)$ for three different sets of light collections optics. Note that 
$C(p)$ first decreases and then oscillates about a baseline intensity. We
arbitrarily define the position of the first minimum to be the average
speckle size. Hence the speckle size is defined as the distance over which
the intensity profile remains spatially correlated.

\begin{figure}[tbp]
\begin{center}
\includegraphics[width=8cm,clip]{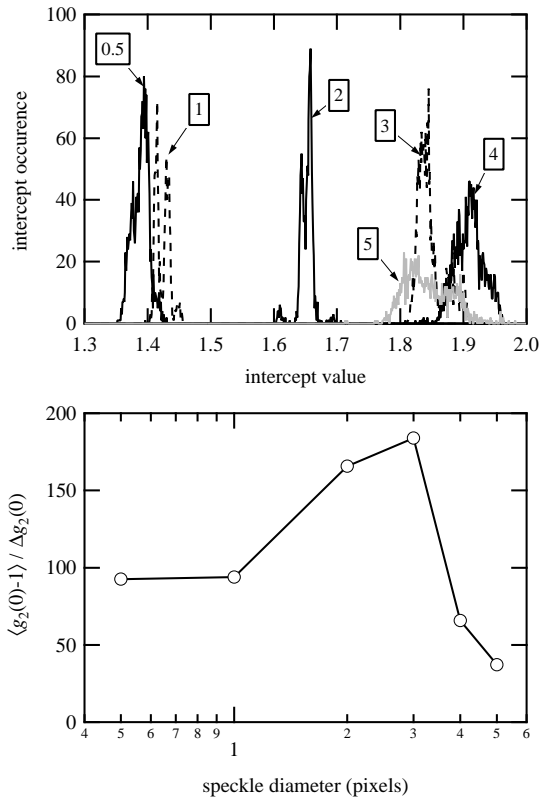}
\end{center}
\caption{(a) Histogram of the intercepts ($g_2(t)=0$) for 2000 correlation
functions for a frit at different speckle size. The theoretical intercept
should be 2. (b) ``Signal-to-noise" $\langle g_2(0)-1\rangle/\Delta g(0)$ as a function of
speckle size where $\langle g_2(0)\rangle$ is the average value of $g_2(0)$
and $\Delta g(0)$ its spread. It shows that the best compromise between lost
of statistics and lost of contrast is obtained for a speckle diameter of 3
pixels.}
\label{fig:statpix}
\end{figure}

To determine optimal configuration of the collection optics, we use the
static scattering pattern from an immobile glass frit. The frit is
sufficiently opaque that diffusion approximation for light transport through
the medium is valid. The intercept for the intensity correlation function (%
\textit{i.e.}\ $g_{2}(0)$) is expected to lie between 1 and 2. In order to
quantitatively evaluate its value, a series of 2000 intercepts was measured.
Measurements were performed by self-correlating 2000 images from the glass
frit. For each image the mean intensity was calculated as well as the mean
squared intensity. The ratio of theses two quantities corresponds to $%
g_{2}(0)$ for this image. Six series of such measurements were made, each
for a different speckle size. Figure \ref{fig:statpix}(a) shows the six
distributions of $g_{2}(0)$. The average value $\langle g_{2}(0)\rangle $
for each speckle size is an evaluation of the contrast, whereas the spread $%
\Delta g_{2}(0)$ indicates how quickly (\textit{i.e.}\ how many images must
be correlated before) $g_{2}(0)$ converges to its proper statistical value.
As such, $\Delta g_{2}(0)$ quantifies the statistical and instrumental
noise. To determine the best compromise between obtaining the highest signal
with the lease noise, we plot $\langle g_2(0)-1\rangle/\Delta g(0)$ as
a function of speckle size. Figure \ref{fig:statpix}(b) shows that the best
configuration is obtained for a speckle size of 3 pixels in diameter. All
subsequent measurements are performed in this configuration. The average
intercept $\langle g_{2}(0)\rangle $ allows the determination of the
normalization factor given for the Siegert relation: 
\begin{equation}
g_{2}(t)-1=\beta |g_{1}(t)|^{2}  \label{eq:siegert}
\end{equation}
where $g_{1}(t)=\langle E(t_{0})E(t_{0}+t)\rangle /\langle
|E(t_{0})|^{2}\rangle $ is the normalized electric field autocorrelation
function. We take $\beta =0.82\pm 0.05$ in subsequent measurements and
display our results in terms of the squared electric field correlation
function $|g_{1}(t)|^{2}$ using the Siegert relation above.

\section{Calculation algorithm}

In this section we explain how the bitmapped image is processed to calculate
the autocorrelation function.

\subsection{Correlation Function Calculation}

At time $t$, a single CCD frame of scattered light is saved in the computer
memory (RAM). We call $I_{i}^{raw}(t)$ the raw intensity of the light
incident on pixel $i$ for the image taken at time $t$, and subtract off an
average background $\langle I_{i}^{back}\rangle $ for pixel $i$ due to stray
light and dark counts: 
\begin{equation}
I_{i}(t)=I_{i}^{raw}(t)-\langle I_{i}^{back}\rangle \;.
\label{eq:intensitydef}
\end{equation}
We measure $\langle I_{i}^{back}\rangle $ by averaging 1000 images for each
pixel with the laser turned off. Further refinements taking into account
other sources of extraneous signal can be found in\ \cite{cipelletti99}.
While such refinements prove useful for DLS in the single scattering limit,
we found that such corrections to be insignificant for DWS.

 One practical problem encountered in acquiring data is that some images may be skipped by the acquisition card due to synchronization problems.
 Thus, as a precaution, we always record the time $t$ at which each image is obtained.

Applying Eq.\ (\ref{eq:intensitydef}) to our data, we calculate the
correlation function: 
\begin{equation}
g_2^{(m)}(t_0,t) = \frac{\langle I_i(t_0) I_i(t+t_0)\rangle_i } {\langle
I_i(t_0)\rangle_i \langle I_i(t+t_0)\rangle}_i
\end{equation}
Note that the average is performed over \emph{pixels} rather than using the
customary method of averaging over the initial time $t_0$. That is, the
correlation function $g_2^{(m)}(t_0,t)$ is calculated by multiplying two
frames (one at time $t_0$, the other at $t_0+t$) together pixel by pixel,
summing the results from all pixels, and then dividing by the number of
pixels. In general, \emph{one frame suffices to obtain a good average} for $%
g_2^{(m)}(t_0,t)$ because it consists of a large number of pixels, $%
16384=128^2$ in this case, which constitutes several thousand independent
samples from which to determine $g_2^{(m)}(t_0,t)$.

\begin{figure}[tbp]
\begin{center}
\includegraphics[width=8cm,clip]{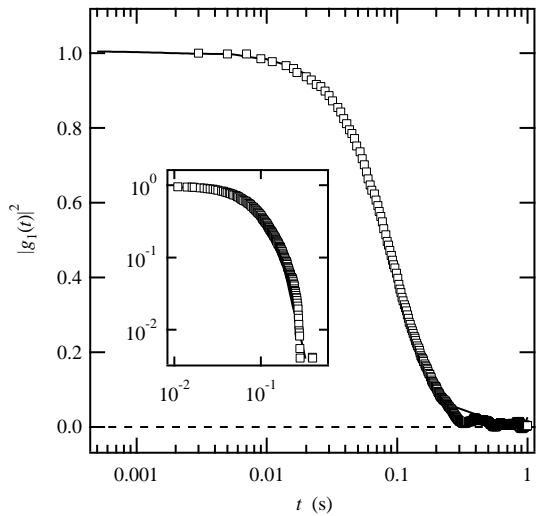}
\end{center}
\caption{Comparison of $|g_1(t)|^2$ obtained from a sample of $0.5~\mu$%
m-diameter beads in pure glycerol for standard single-correlator DWS set up (%
$-$) and multi-speckle camera based set up (Square). The inert shows the
correlation function with the baseline subtracted demonstrating that
reliable data is obtained down to 10$^{-2}$.}
\label{fig:cameraPM}
\end{figure}

The correlation function $g_2^{(m)}(t_0,t)$ is obtained in this manner is
obtained from an \emph{ensemble average} of the sample, with no need for the
time averaging method typically used with conventional light detection. When
data is taken from an ergodic system, both methods give the same result, as
illustrated in Fig.\ \ref{fig:cameraPM} where we compare correlation
functions obtained from a CCD camera to one taken with a conventional single
detector-single correlator for a sample of polystyrene beads in glycerol
obtained undergoing Brownian motion. Significant advantages accrue for the
CCD technique, however, when measurements are performed on a non-ergodic
sample. Moreover, the CCD method can provide important advantages when
measuring correlation functions for non-ergodic systems.

\subsection{Multiple correlation function calculation}

In the preceding section, we discussed how a two-time correlation function $%
g_2(t_0,t)$ can be obtained, $t_0$ being the time at which the reference
image is acquired and $t_0+t$ being a later time at which another image is
collected and correlated with the reference. Such a correlation function
provides a snapshot of the dynamical state of the system at a particular
starting time, namely $t_0$, and is possible only because sufficient
ensemble averaging is achieved by correlating thousands of pixels between
two frames from a CCD camera. The ability to obtain such snapshots is
particularly useful for studying systems whose dynamics are time-dependent, 
\textit{i.e.}\ non-stationary systems. Examples include systems which are
gelling, coarsening, aging, or undergoing some other non-equilibrium
non-stationary process. By measuring $g_2(t_0,t)$ as a function of the
starting time $t_0$, the evolving dynamics of such systems can be followed
and studied.

Here, we explain a scheme for following the evolving dynamics of a system by
measuring many different correlation functions in parallel, each with a
different starting time $t_0$. A unique and important feature of this scheme
is that new correlation functions can be started at intervals as short as
1/500~s, while still collecting frames for correlation functions started at
earlier times. In fact, up to 500 correlation functions can be collected
simultaneously, each with a different starting time $t_0$, and without any
requirement that a previously started correlation function be finished
before starting a new one.

The key to the success of the scheme is the realization that in most cases,
it is not necessary to measure the correlation function with the same
temporal resolution for all delay times. Typically, small time steps are
required for short delay times while larger time steps are required for
longer delay times. The scheme we employ takes advantage of this fact. It is
essentially an interlacing scheme in which successive frames are sometimes
used in different correlation functions, as will be explained below. In
effect, the scheme represents a compromise between making maximal use of the
information available in the measurements for calculating the correlation
functions and obtaining and processing the data from the CCD camera frames
at a rapid rate.

This compromise is necessary because calculating the correlation function
between two frames is time consuming; only one pair of frames can be
processed (multiplied and averaged) between the acquisition of two
successive images. Image stacking is not practical because it quickly leads
to memory overflow. Therefore, we employ an algorithm for calculating many
correlation functions, labeled by $j=1,2,3,\ldots, j_{max}$, simultaneously
in real time, that makes only the following manageable demands on the
computer:
\begin{itemize}
\item  only one frame multiplication is performed between the acquisition of successive frames,
\item  storage of a maximum of $j_{max}\lesssim 500$ reference image frames
plus the most recently acquired frame in memory, and
\item  storage of a maximum of $j_{max}$ correlation functions in memory,
each with on the order of 100 channels.
\end{itemize}

\begin{figure*}[tbp]
\begin{center}
\includegraphics[width=14cm,clip]{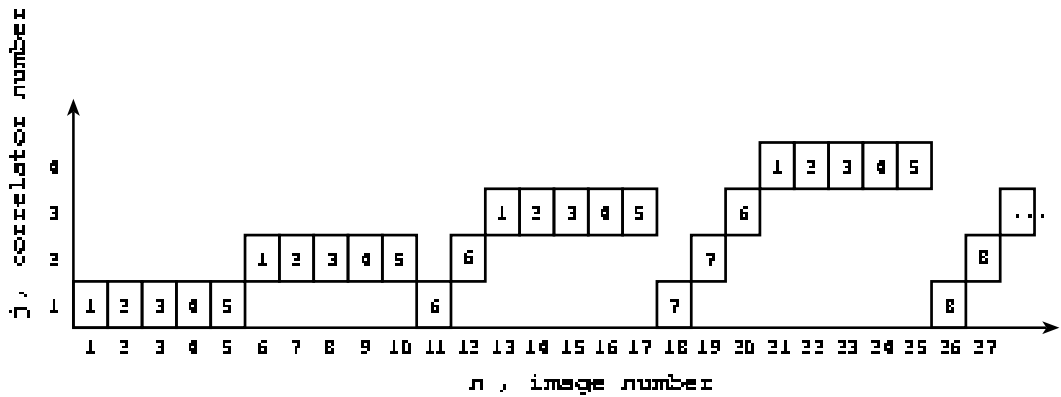}
\end{center}
\caption{Schematic representation of image selection and use in the
multicorrelator scheme. For this illustration $N=5$. The horizontal axis is
the image number $n$ (or $n_i^{(j)}$) and the vertical axis is the
correlator number $j$. The numbers inside the boxes indicate the $i^{th}$
image used in correlator $j$.}
\label{fig:multicorrprinc}
\end{figure*}

For a frame rate of 500 frames/s from a $128\times 128$ pixel CCD camera,
these demands can be met by a computer with a Pentium processor running at
500~MHz and possessing 512~MB of RAM.
 An important feature of the algorithm is that it requires that only the reference frame for each correlation function, along with most recently acquired frame, be stored in the computer at any moment in time; this keeps the number of frames that must be stored in memory to a manageable number.
 This also allows all correlation functions to be calculated in real time as the experiment proceeds.

The scheme we use is illustrated graphically in Fig.\ \ref
{fig:multicorrprinc}. The vertical axis represents the different correlation
functions labeled by $j=1,2,3,\ldots, j_{max}$. The horizontal axis
represents the successive frames that are acquired, labeled $n=1,2,3,\ldots$
with each frame being nominally separated in time by the reciprocal $1/f$ of
the frame rate. The reference image for the $j^{th}$ correlation function is
given by image number 
\begin{equation}
n = n_1^{(j)}=1+(N+1)(j-1) \;.
\end{equation}
where the subscript ``1" indicates that this is the first frame of the $%
j^{th}$ correlation function. Each of these reference images is stored in
memory; $N$ is typically of order 10. The reference image and the next $N-1$
images in sequence are correlated with the reference image. The remaining
frames used to calculate the $j^{th}$ consists of a sequence of selected
images as illustrated in Fig.\ \ref{fig:multicorrprinc}. The sequence
numbers of the frames used to calculate the $j^{th}$ correlation function is
summarized by the formula: 
\begin{widetext}
\begin{equation}
n = n_i^{(j)} = \left\{ 
\begin{array}{ll}
n_1^{(j)} + i-1 = (N+1)(j-1) + i & \mbox{~~if~~$1 \le i \le N$} \\ 
(j-1) + \frac{1}{2}i(i-1)-\frac{1}{2}N(N-1) & \mbox{~~if~~$i>N$}
\end{array}
\right. \;.
\end{equation}
\end{widetext}
Note that the spacing between the points in a correlation function is
constant over the first $N$ points but increases quadratically at later
times. The time interval between two successive correlation functions $j$
and $j+1$ is approximately $N/f$ where $f$ is the frame rate. This time
interval can be tuned in order to best accommodate the rate of evolution of
the system under study. For $N=1$ this time interval can be as short as the
inverse frame rate.

\begin{figure}[tbp]
\begin{center}
\includegraphics[width=8cm,clip]{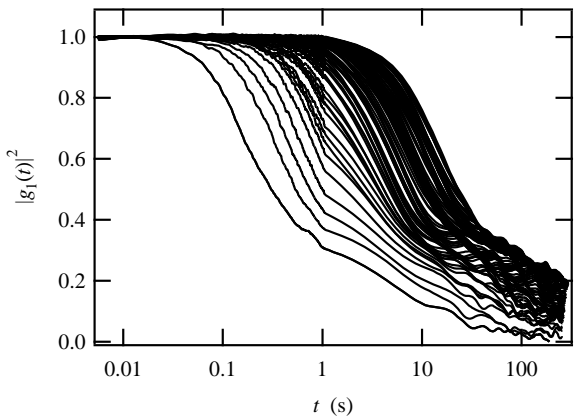}
\end{center}
\caption{Correlation functions $|g_1(t_0,t)|^2$ taken after a shear step of
$5\%$.
 Multiple correlation functions are obtained in real time with the ones started later having larger decorrelation times; the decorrelation time increases as the relaxation process slows down.
 The delay between two successive correlation function is approximately 1 sec.}
\label{fig:relaxation}
\end{figure}

\begin{figure}[tbp]
\begin{center}
\includegraphics[width=8cm,clip]{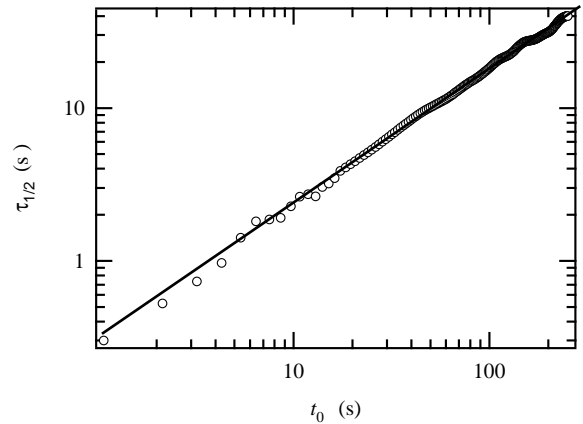}
\end{center}
\caption{The correlation time $\tau_{1/2}$ vs.\ the time $t_0$, where $%
\tau_{1/2}$ is defined as the delay time where $g_{1}(t_0,\tau_{1/2})=\frac{1}{2}$, and $t_0$ is the amount of time that passed after the step strain
before starting the correlation function. A new correlation function was
started every 1~s. The data are well-described by a power law $\tau_{1/2}
\sim t_0^{0.88}$.}
\label{fig:taudemi}
\end{figure}

In Fig.\ \ref{fig:relaxation} we show a series of correlation functions
obtained using the multispeckle CCD camera technique described above. The
correlation functions were obtained from a sample consisting of 0.5 wt.\%
alumina (Al$_{2}$O$_{3}$) beads, approximately 160~nm in diameter (AKP30
Sumitomo), that were embedded in a crosslinked elastic network of
polybutylacrylate. The first correlation function was started at a time $%
t_0=0$ immediately after a step strain of 5\% was applied to the sample. Thus,
the correlation functions measure the subsequent random motion of the beads
that arises as a response to the stress within the sample. As the stress
relaxes over the course of tens of minutes, the random motion of the beads
becomes slower. This slowing down is reflected in the increased
decorrelation time of correlation functions started at increasingly later
times after the initial step strain. In Fig.\ \ref{fig:taudemi} we plot the
decorrelation time $\tau _{1/2}$, defined as the time it takes a correlation
function to decay to half its initial value, vs.\ time, referenced to the
starting time $t_0$. It exhibits a power law behavior over two decades. A
detailed discussion of this experiment will be published elsewhere.

\section{Fast and slow time scales}

\label{sec:fastandslow}

The shortest decorrelation time $\tau_{1/2}$ measured in the experiments
described in the previous section is fairly slow, approximately 0.1~s. In
this particular instance, the frame rate of the camera is sufficient to
capture all the dynamics of the process in the correlation function. This is
not always the case, however. Many materials, including polymer networks,
colloidal gels, and glassy materials possess two characteristic ranges of
relaxation times, where only the slower of the two is sufficiently slow for
the multispeckle CCD camera method to work. In these cases, the faster
relaxation processes are not measured.

\begin{figure}[tbp]
\begin{center}
\includegraphics[width=8cm,clip]{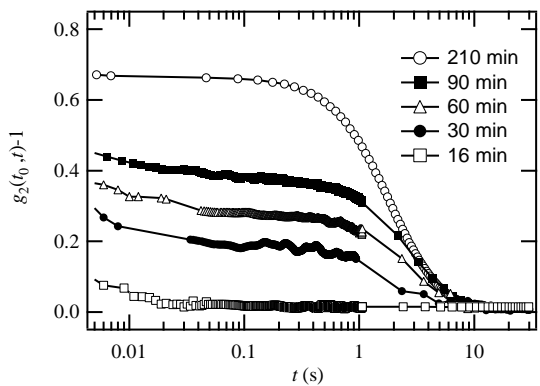}
\end{center}
\caption{Correlation functions $g_2(t_0,t)-1$ ($=\beta|g_1(t_0,t)|^2$)
started at different times $t_0$ (indicated in plot) after the quench of a
1\% gelatine gel with 1\% latex spheres from $60^\circ$C to $20^\circ$C. The
typical evolution time is around 10 minutes. The rise of the intercept and
plateau corresponds to the disappearance of the fast liquid phase relaxation
process and concurrent increase of the gel elastic modulus.}
\label{fig:campart}
\end{figure}

To illustrate such a process, we use the multispeckle technique to follow
the gelation that occurs following a quench from $60^\circ$C to $20^\circ$C in a 1\% gelatin solution containing 1\% latex spheres with a radius of
85~nm. The correlation functions are shown in Fig.\ \ref{fig:campart}. As
the gelatin gels, the motion of the strands of gelatin and the motion of the
spheres embedded within it becomes increasingly restricted. As the gelation
proceeds, the height of the plateau and the $y$-intercept of the correlation
functions increases. This occurs because there is a fast relaxation process
that occurs outside of the time window of the camera. That is, there is an
initial rapid drop of $g_{2}(t)$ that occurs on a time scale that is faster
than the inverse frame rate of the camera ($\sim 2$~ms). Below, we explain
how this obstacle can be overcome such that the lower bound of the time
window is reduced from $10^{-2}$ to $10^{-8}$~s.

What we would like to be able to do is to \emph{simultaneously} capture the
fast and slow dynamics for a time-dependent non-stationary process. Below,
we show that this can be accomplished, provided at strict set of criteria
are met. Fortunately, it turns out that these criteria are met by most of
the systems of interest so that the criteria are not very restrictive
in practical situations. As we will show below, we can use a classical
photomultiplier-based (fast) correlator, which spans a delay-time range of $%
10^{-8}$~s to about $1$~s, to capture the fast dynamics and the multispeckle
(slow) correlator to capture the slow dynamics. The key is to properly
ensemble average the scattered light signal entering the photomultiplier
used with the fast correlator without disturbing the multispeckle signal
that enters the slow correlator.

To provide an ensemble-averaged signal to the fast correlator, we exploit a
technique recently developed by Scheffold \textit{et al.}~\cite{romer00},
which enables simple and efficient spatial averaging without requiring any
motion of the sample or the detector. This technique is known as the \textit{%
two-cell} technique and has been described both theoretically \cite{skipetrov96} and experimentally \cite{romer00}.

\subsection{Two-cell technique}

The idea of the two-cell technique is to optically couple the studied sample
to an ergodic reference system. The light is successively scattered by the
sample cell and then by the reference cell before entering the photomultiplier.
 This can be implemented simply by illuminating two cells in
series, and detecting the light that passes through and is scattered by both
samples in the two cells. In that case, the non-ergodic signal emerging from
the sample cell is made ergodic by its scattering through the reference
cell. The second cell can thus be regarded as performing a slow spatial
averaging of the sample signal. Under carefully specified optical
conditions, the correlation function, $g_{1}^{M}(t),$ measured after passing
through the two cells is simply the product of the correlation function of
each cell: 
\begin{equation}
g_{1}^{M}(t)=g_{1}^{S}(t)\times g_{1}^{R}(t)\;,  \label{eq:g2cell}
\end{equation}
where the superscripts $R$ and $S$ stand for the sample and the reference
correlation functions, respectively. The conditions under which the two-cell
technique can be employed have been carefully considered in Ref.\ \cite{skipetrov96}. 
They find that the primary condition required is that each
photon traverses the space between the sample and reference cells only once.
This means that for a configuration where a significative amount of light is lost between the two cells, as was done in Ref.\ \cite{romer00}%
, this condition is satisfied when $l_{R}^{*}/L_{R}\gg l_{S}^{*}/L_{S}$,
where $l_{i}^{*}$ and $L_{i}$ are, respectively, the photon transport mean
free path and the cell thickness for the reference ($i=R$) and the sample ($%
i=S$) cells.

\begin{figure*}[tbp]
\begin{center}
\includegraphics[width=12cm,clip]{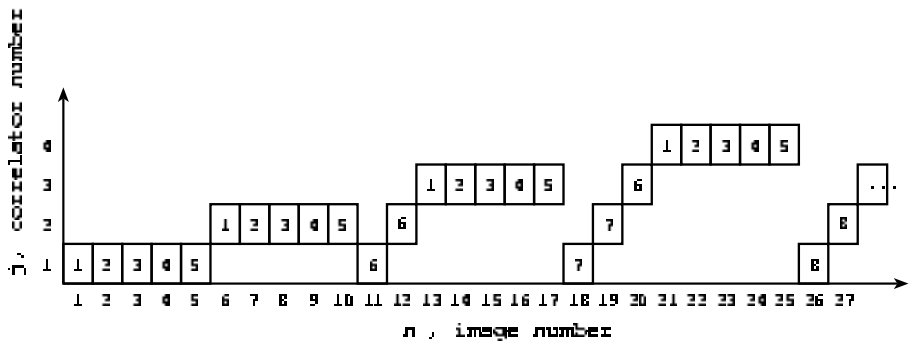}
\end{center}
\caption{Schematic for the two-cell technique. A single mode optic fiber and
a multi-delay-time correlator with a photomultiplier tube are used. The
second cell is made with $0.5~\mu$m polystyrene spheres in a dilute solution
of glycerol. The polarizing cube has been replaced by a non-polarizing beam
splitter.}
\label{fig:dessinmontage2}
\end{figure*}

\subsection{Experimental set up}

Figure \ref{fig:dessinmontage2} illustrates how we incorporate the two-cell
technique into the multispeckle correlator so that we can simultaneously
capture the fast and slow dynamics from a slowly evolving non-ergodic
system. We arrange the optics for the multispeckle (slow) correlator just as
before so that surface from which the multiply scattered light exits the
sample cell is imaged with a 1:1 ratio onto a pinhole behind which the CCD
camera is placed. For the fast correlator, a non-polarizing beam splitter is
placed a short distance in front of the pinhole so that  half of the light
is diverted onto the reference cell. The reference cell positioned so that
the exit plane of the sample cell is imaged onto the reference cell's front
surface. An single mode optical fiber with a collimator collects the
scattered light exiting the reference cell at an angle of approximately $%
15^{\circ }$ and sends it to a photomultiplier whose output is connected to
a multiple-sample-time correlator.

The reference cell has a thickness of 5~mm and contains a glycerol solution
of 0.5~$\mu$m-radius latex spheres with a volume fraction of $1\%$. The
concentration of spheres is chosen so that there is just enough multiple
scattering in the reference cell to ensure that essentially no light passes
through the sample without being scattered at least several times but with $l_{R}^{*}/L_{R}$ as big as possible. However,
this must be done so that the decorrelation time of the reference
correlation function is somewhat larger, typically by a factor of about 3 to
10, than the inverse frame rate of the CCD camera. By adjusting the
concentration of spheres and the viscosity of the solution with glycerol,
these conditions are readily achieved.

\begin{figure}[tbp]
\begin{center}
\includegraphics[width=8cm,clip]{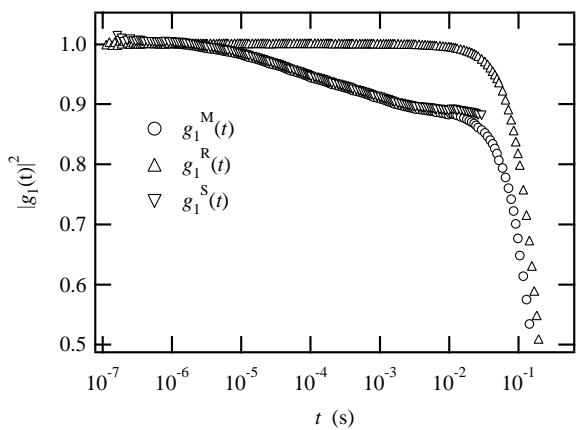}
\end{center}
\caption{Correlation functions $|g_{1}^M(t)|^2$ [$\circ$], $|g_{1}^R(t)|^2$ [%
$\bigtriangleup$] $|g_{1}^S(t)|^2$ [$\bigtriangledown$] for the gelatine gel
after 210 min. The sample correlation function $|g_{1}^S(t)|^2$ displays
first a rapid decay and then a plateau due to the trapping of the beads in
the gel network.}
\label{fig:proof}
\end{figure}

Before a measurement can be made using the two-cell technique, it is
necessary to measure the correlation function of the reference cell in the
absence of the sample cell and to ensure that its correlation time is
greater than the inverse frame rate. Data from such a measurement is shown
in Fig.\ \ref{fig:proof}, from which one can see that the decay time of $%
g_1^R(t)$ is about 0.1~s. The correlation function $g_1^M(t)$ is then
measured with the sample cell in place, from which the sample correlation
function $g_1^S(t)$ is obtained using Eq.\ (\ref{eq:g2cell}). Figure \ref
{fig:proof} shows results from a typical set of measurements, including the
reference cell correlation function $g_1^R(t)$, the measured two-cell
correlation function $g_1^M(t)$, and the true correlation function of the
sample $g_1^S(t)$ obtained from Eq.\ (\ref{eq:g2cell}). Note that $g_1^S(t)$
is obtained accurately from the first time measured by the correlator, about 
$10^{-7}$~s, out to about 30~ms, or approximately the decorrelation time of $%
g_1^R(t)$. For longer delay times, the reference correlation function $%
g_1^R(t)$ falls to zero making recovery of $g_1^S(t)$ using Eq.\ (\ref
{eq:g2cell}) impossible. Note that the sample correlation function $g_1^S(t) 
$ determined in this way has a significant time interval of overlap, from
about 2~ms to about 30~ms, with the correlation function measured by the
multispeckle technique. Thus, it should be possible to superimpose the
correlation functions from the two techniques to obtain a single correlation
function covering the entire range of time scales from 10~ns to hundreds of
seconds or longer. It is important to note, however, that in order from the
signal to be ergodic the correlation function obtained from the two-cell
technique to be properly ensemble averaged, data must be collected over a
time that is about $10^3$ times larger than the decay time of the reference
correlation function. In our case the acquisition time is 100~s.

\subsection{Overlapping data from the two techniques}

In order to overlay the correlation functions measured by the two
techniques, they must be properly normalized using the respective Siegert
coefficients $\beta$ for each measurement (see Eq.\ \ref{eq:siegert}). For
the multispeckle CCD camera method, $\beta=0.8$, as discussed in Section \ref
{sec:spklsz}. For the double cell technique, $\beta$ is determined by the
value of the intercept from the correlation function of the reference cell
alone. For the reference cell, we obtain $\beta=0.95$.

\begin{figure}[tbp]
\begin{center}
\includegraphics[width=8cm,clip]{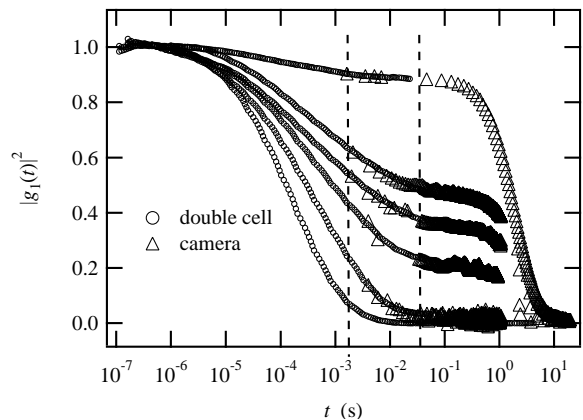}
\end{center}
\caption{Normalized correlation functions obtained using the double cell ($%
\circ$) and multispeckle ($\bigtriangleup$) techniques. The range of overlap
is indicated by the dashed lines.}
\label{fig:overlay}
\end{figure}

Figure \ref{fig:overlay} shows correlation functions from the two techniques
for the gelation experiment described in Section \ref{sec:fastandslow}. The
acquisition time for each double cell measurement was 100~s. Excellent
agreement is obtained over the range where the two techniques overlap
(2~ms-30~ms). Note that the overlap occurs in a region where the slope of
the correlation function is sizeable, making the agreement even more
impressive.

\begin{figure}[tbp]
\begin{center}
\includegraphics[width=8cm,clip]{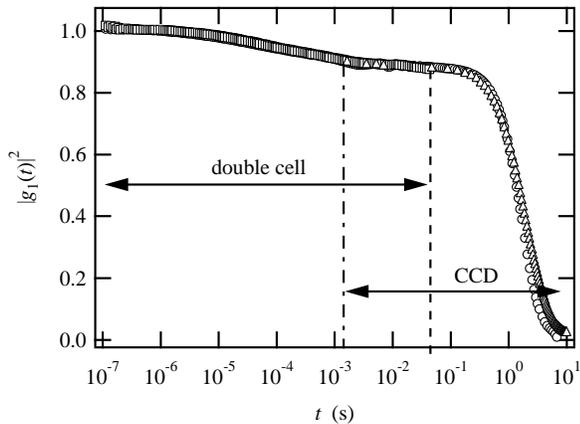}
\end{center}
\caption{Comparison of correlation functions obtained using a single
correlator without a reference cell (conventional single-cell method --
3000~s) [$\bigcirc$], with a reference cell (two-cell method -- 100~s) [Square], and using multispeckle CCD camera (10~s) [$\triangle$] techniques.
Discrepancies at long time at are attributed to poor statistics for the the
conventional method.}
\label{fig:verif}
\end{figure}

Finally, it us useful to compare the results for different methods of
measuring a correlation function. In Fig.\ \ref{fig:verif}, we plot three
correlation functions: (1) one obtained using the multispeckle technique
with a window of decay times of 2~ms--10~s. Averaging was performed in real
time (up to the longest decay time or $\sim 20$~s); (2) another obtained
using the double cell technique over the decay-time window of $10^{-7}$%
s--30~ms and averaged for 100~s; (3) and a third obtained using a classical
set-up with a photomultiplier tube and single correlator over the decay-time
window of $10^{-7}-5$~s and averaged for 3000~s. The measurements using the
double cell and the multispeckle methods were done simultaneously; the
measurement with the photomultiplier and single correlator was performed
just after the other two, simply by removing the reference cell. The
agreement is very good over the time window of $10^{-7}$-$10^{-1}$~s. There
are deviations, however, between the single correlator and the multispeckle
techniques at longer delay times. These deviations occur because data
obtained using the single correlator technique must be collected over a
fairly long data-acquisition time, 100~s in this case, in order to obtain
satisfactory signal-to-noise. In cases where the dynamics of the sample
evolves with time, the multispeckle technique has clear advantages.

\begin{figure}[tbp]
\begin{center}
\includegraphics[width=8cm,clip]{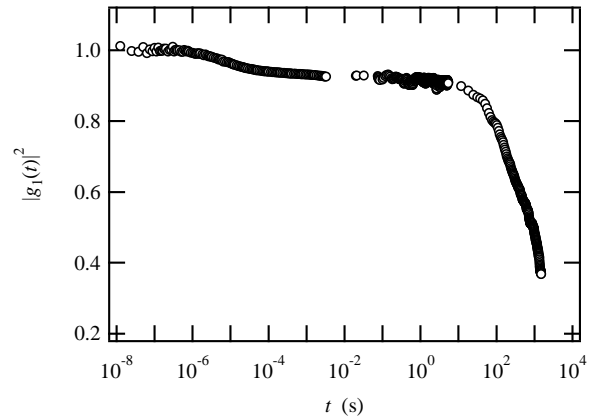}
\end{center}
\caption{Correlation function $|g_1(t)|^2$ for a suspension of silica
spheres ($\phi =22\%$). The fast part of the correlation function was
measured using the two-cell technique and the slow part using the
multispeckle technique. The combined measurements span 11 decades.}
\label{fig:silice}
\end{figure}

The multi-speckle technique can measure correlation functions in real time
for delay-times larger than 2~ms. When coupled simultaneously (in the same
set-up) with the two-cell technique, the combined method can measure
correlation functions over an extremely broad range of delay times. As a
demonstration, we show in Fig.\ \ref{fig:silice} a correlation function
measured for a gel of silica beads (0.2 $\mu m$ in diameter) at a volume
fraction of 52\% with an Ionic force adjusted to $1.4$ $10^-2$M with KNO$_3$. correlation function spans a range of delay times from
10$^{-8}s$ to greater than 10$^3$~s, or more than 11 decades in time. Such a
measurement is not only remarkable, it is extremely useful for probing
glassy systems which can exhibit dynamics over a vary broad range of delay
times.

Lastly, there are a couple of practical constraints that bear mentioning.
First, the two techniques can be usefully coupled only if the typical time
scale over which the system evolves is much longer than the averaging time
required for the two cell technique. As a practical matter, the required
averaging time is set by the longest delay time one wishes to measure. That
time is determined primarily by the desire to have some range of delay times
where the two-cell and multi-speckle techniques overlap. The limitation here
is the speed of the camera that can be used for the multi-speckle technique.
Second, the simultaneous use of the two techniques leads to a certain amount
of light loss. Thus, a powerful laser ($\sim 1$~W) is recommended.



\bibliographystyle{unsrt}

\end{document}